\documentstyle[preprint,aps]{revtex}

\input psfig.sty

\begin{document}
\draft
\title{ Neutrino emission due to Cooper pairing of protons in cooling neutron 
stars: Collective effects }
\author{L. B. Leinson\footnote{On leave from:
Institute of Terrestrial Magnetism, Ionosphere and Radio Wave 
Propagation RAS \\
142092 Troitsk, Moscow Region, Russia}}
\address{Departamento de F\'{i}sica Te\'{o}rica, Universidad de Valencia\\
46100 Burjassot (Valencia), Spain.}
\maketitle

\begin{abstract}
The process of neutrino-pair radiation due to formation and
breaking of Cooper pairs of protons in superconducting cores of neutron
stars is considered with taking into account of the electromagnetic coupling of
protons to ambient electrons. It is shown that plasma polarization strongly
modifies the effective vector weak current of protons. Collective response
of ambient electrons to the proton quantum transition contributes coherently
to the complete interaction with the neutrino field and enhances the rate of
neutrino-pair production by two orders of magnitude. 
\end{abstract}
\widetext
Except for the early stages of evolution, temperatures inside neutron stars
are lower than the critical temperatures for proton superconductivity 
$T_{cp} $ and/or neutron superfluidity $T_{cn}$. Conventionally, neutron
superfluidity as well as proton superconductivity, exponentially reduce the
heat capacity of neutron matter and the rates of neutrino production from
nucleon reactions in the core of a neutron star \cite{Tsur72}, \cite{Page92}
, \cite{Page94}, \cite{Leven96}. When $T<T_{cp},T_{cn}$, nuclear matter in
the stellar core consists of two components. One part of nucleons forms a
condensate of Cooper pairs, the other one represents unpaired
quasi-particles. Both processes, Cooper-pair formation and pair-breaking
coexist in statistical equilibrium and result in additional neutrino-pair
emission from the neutron star. This mechanism of energy loss from
superfluid neutron stars was first proposed and evaluated for neutron
singlet-state superfluidity by Flowers et al. \cite{FRS76}, but it has been
included into cooling simulations only recently \cite{Vosk97}, \cite{Page98}
, \cite{Yak98}. It was shown, that the phase transition of neutrons in a
superfluid state does not decelerate the process of cooling, but, on the
contrary, neutrino emission due to formation and breaking of Cooper pairs
accelerates the neutron star cooling.

   Cooper pairing of protons take place likely in $^{1}S_{0}$\ 
state \cite{TT93}. Since the total spin of the Cooper pair in the $^{1}S_{0}$ 
state is zero,
the axial-vector contribution of the weak neutral current vanishes.
Contribution to $\nu \bar{\nu}$ emissivity comes only from the vector weak
interaction. Therefore, neutrino emission produced by proton pairing is
conventionally assumed to be strongly suppressed due to the numerical
smallness of the weak vector coupling constant of protons \cite{YKL98}. Such
inference is made on the basis of calculations which ignored electromagnetic
correlations among the charged particles in a QED plasma. Actually, protons
in the plasma are coupled to ambient electrons via the electromagnetic
field. By undergoing a quantum transition to the paired state, protons
polarize the medium, thus inducing the motion of electrons inside the Debye
sphere around them. The electron weak current associated to this motion
generates neutrinos coherently with the weak current of protons, because the
wavelength $\lambda $ of radiated neutrino pairs is much larger than the
electron Debye screening distance $D_{e}$ (typically, $D^2_{e}/\lambda^2
\lesssim 10^{-2}$). In the present article we study an effective weak
current of protons with taking into account the collective contribution of
ambient electrons, and estimate the neutrino-pair emissivity due to
formation and breaking of proton Cooper pairs.

   In what follows we use the system of units $\hbar =c=1$ and the Boltzmann
constant $k_{B}=1$. The low-energy Lagrangian of proton interaction with the
neutrino field in vacuum can be written in a point-like approach 
\begin{equation}
{\cal L}_{vac} =-\frac{G_{F}}{\sqrt{2}}\bar{\nu}\gamma _{\mu }\left(
1-\gamma _{5}\right) \nu j^{\mu },  \label{H}
\end{equation}
where $G_{F}$ is the Fermi coupling constant, and the proton weak current is
of the form 
\begin{equation}
j^{\mu }=\bar{\psi}\gamma ^{\mu }(C_{V}-C_{A}\gamma _{5})\psi  \label{jp}
\end{equation}
Here $\psi $ stands for the proton field; $C_{V}$ and $C_{A}$ \ are the
vector and axial-vector coupling constants of the proton, respectively.
Reserving the capital letter notations $C_{V}$ and $C_{A}$ \ for proton
coupling constants, we will use, at the same time, small letters, $c_{V},$
for electron coupling constants with $c_{V}=\frac{1}{2}+2\sin ^{2}\theta
_{W}\simeq 0.96$ for electron neutrinos, and $c_{V}^{\prime }=-\frac{1}{2}
+2\sin ^{2}\theta _{W}\simeq -0.04$ \ for muon and tau neutrinos; $\theta
_{W}$ is the Weinberg angle.

   To describe the above-mentioned collective effects, one has to include in
the matrix element of the reaction two Feynman diagrams shown in Fig. 1. The
sum of these diagrams represents the effective weak coupling of in-medium
protons with the neutrino field. The matrix element of this interaction
should be calculated between exact initial and final states of two proton
quasi-particles undergoing quantum transition to the paired state. The first
diagram in Fig. 1 represents the proton weak interaction in vacuum, and the
second one is the contribution of ambient electrons. Some explanations are
necessary here. At the first sight, the second diagram contains additionally
a small factor $e^{2}=1/137$. However, the fine structure constant enters
the matrix element in a combination with the electron number density, known
as the Debye screening distance. Therefore, the parameter of the problem is $
k^{2}D_{e}^{2}$, where $k$ is the momentum carried out by the neutrino-pair.
As will be shown, in the case $k^{2}D_{e}^{2}\lesssim 1$ the second diagram
gives an important contribution and should be necessarily included in the
matrix element of the reaction.

   The electron loop in the second diagram can be expressed in terms of
polarization tensors of the electron gas, defined in the one-loop
approximation 
\begin{equation}
\Pi ^{\,\mu \rho }\left( K\right) =4\pi ie^{2}{\rm Tr}\int \frac{
d^{4}p}{(2\pi )^{4}}\,\gamma ^{\mu }\,{\hat{G}}(p)\,\gamma ^{\rho }\,{\hat{G}
}(p+K),  \label{PT}
\end{equation}
\begin{equation}
\Pi _{5}^{\,\mu \rho }\left( K\right) =4\pi ie^{2}{\rm Tr}\int \frac{
d^{4}p}{(2\pi )^{4}}\,\gamma ^{\mu }\,{\hat{G}}(p)\,\gamma ^{\rho }\,\gamma
_{5}{\hat{G}}(p+K).  \label{PA}
\end{equation}
Here ${\hat{G}}(p)$ is the in-medium electron Green's function, and $
K=\left( \omega ,{\bf k}\right) $ is the four-momentum transfer. Thus,
the in-medium effective weak current of protons is of the following form 
\begin{equation}
J^{\mu }=\bar{\psi}\gamma ^{\mu }(C_{V}-C_{A}\gamma _{5})\psi -\frac{1}{4\pi 
}\left( \bar{\psi}\gamma ^{\lambda }\psi \right) D_{\lambda \rho }\left(
c_{V}\Pi ^{\,\rho \mu }-c_{A}\Pi _{5}^{\,\rho \mu }\right) ,  \label{Jp}
\end{equation}
where $D_{\lambda \rho }\left( K\right) $ is the photon propagator in the
medium. An extra factor $\left( 4\pi \right) ^{-1}$ appears here because the
factor $4\pi $ is traditionally included in the definition of polarization
tensors. The minus sign in front of this factor is due to the fact that the
second diagram includes a proton charge $e>0$ in the left-hand
electromagnetic vertex of the virtual photon, and an electron charge $-e$ in
the right-hand vertex. The corresponding $e^{2}$ factor which appears
accordingly to the second diagram, is also included in the definition of
polarization tensors.

   To write the in-medium photon propagator one has to take into account that
the superconducting electric current of protons and the normal electric
current of electrons coexist, participating in electromagnetic oscillations
of the medium. The total Lagrangian density of the complex field $\Psi $ of
proton Cooper pairs, and electrons interacting with the electromagnetic
field $A_{\mu }$ has the form 
\begin{equation}
L=\left| \left( \partial _{\mu }+2ieA_{\mu }\right) \Psi \right|
^{2}+M_{Cp}^{2}\left| \Psi \right| ^{2}-\lambda \left| \Psi \right| ^{4}-
\frac{1}{16\pi }F_{\mu \nu }^{2}-j_{\mu }^{e}A^{\mu }+L_{e}^{0}
\label{Ldens}
\end{equation}
where $\lambda \ $is a constant for proton-proton interaction resulting in
the proton pairing, and $M_{Cp}\simeq 2M_{p}^{\ast }$ is the mass of the
Cooper pair consisting of two protons of effective mass $M_{p}^{\ast }$; $
F_{\mu \nu }=\partial _{\mu }A_{\nu }-\partial _{\nu }A_{\mu }$ is the
tensor of electromagnetic field; $j_{e}^{\mu }$ is the electron current, and 
$L_{e}^{0}$ is the Lagrangian density of free electrons. As a result of
spontaneous breaking of the symmetry of the ground state of the system, the
vacuum expectation value of the Cooper pair field is nonzero $\left\langle
\left| \Psi \right| ^{2}\right\rangle $ =$\Psi _{0}^{2}$. Since the ground
state of the system corresponds to zero spin and zero total momentum of the
Cooper pair the vacuum expectation value is connected with the number
density of Cooper pairs $N_{s}$ by the relation $N_{s}=2E\Psi _{0}^{2}$,
where $E\approx 2M_{p}^{\ast }$ is the energy of the Cooper pair
corresponding to zero total momentum, and $N_{s}$ is one-half the number
density of paired protons $N_{p}$. Thus, 
\begin{equation}
\Psi _{0}^{2}=\frac{N_{p}}{8M_{p}^{\ast }}.
\end{equation}
We represent the field of the Cooper condensate in the form 
\begin{equation}
\Psi =\Psi _{0}\left( 1+\rho \right) \exp \left( -2ie\phi \right) 
\end{equation}
where $\rho $ and $\phi $ are arbitrary real functions of the coordinates
and time, and subsitute in the Lagrangian density (\ref{Ldens}) for the
fields $\rho $ and $\phi $. Taking into account the following identity 
\begin{equation}
\left| \left( \partial _{\mu }+2ieA_{\mu }\right) \Psi \right| ^{2}=\Psi
_{0}^{2}\left( \partial _{\mu }\rho \right) ^{2}+4e^{2}\Psi _{0}^{2}\left(
A_{\mu }-\partial _{\mu }\phi \right) ^{2}\left( 1+\rho \right) ^{2}
\end{equation}
we can make the gauge transformation $A_{\mu }^{\prime }=A_{\mu }-\partial
_{\mu }\phi $. Since the quantity $F_{\mu \nu }=F_{\mu \nu }^{\prime
}=\partial _{\mu }A_{\nu }^{\prime }-\partial _{\nu }A_{\mu }^{\prime }$ as
well as the electron current $j_{\mu }^{e}=j_{\mu }^{\prime e}$ are
gauge-invariant we obtain
\begin{equation}
L=\psi _{0}^{2}\left( \partial _{\mu }\rho \right) ^{2}+4e^{2}\Psi
_{0}^{2}A_{\mu }^{\prime 2}\left( 1+\rho \right) ^{2}+M_{Cp}^{2}\left| \psi
\right| ^{2}-\lambda \left| \psi \right| ^{4}-\frac{1}{16\pi }F_{\mu \nu
}^{\prime 2}-j_{\mu }^{e}A^{\prime \mu }+L_{e}^{0}
\end{equation}
As a result, the Goldstone field $\phi $ is absorbed by the gauge
transformation. Considering this and taking variations of the Lagrangian
with respect to the field $A^{\prime \mu }$, in the linear approximation we
obtain the equation 
\begin{equation}
\partial _{\nu }\partial ^{\nu }A_{\mu }^{\prime }+32\pi e^{2}\Psi
_{0}^{2}A_{\mu }^{\prime }=4\pi j_{\mu }^{e},\ \ \ \ \partial ^{\mu
}A_{\mu }^{\prime }=0  \label{A}
\end{equation}
In the absence of the electron current $\left( j_{\mu }^{e}=0\right) $, this
equation would describe the eigen photon modes of mass $m_{\gamma }=\sqrt{
32\pi e^{2}\Psi _{0}^{2}}=\sqrt{4\pi N_{p}e^{2}/M_{p}^{\ast }}$. This is the
well-known Higgs effect. However, the induced electron current should also
be taken into account. In the momentum representation $4\pi j_{\mu
}^{e}=-\Pi _{\mu \nu }A^{\prime \nu }$, and the Eq.(\ref{A}) takes the form 
\begin{equation}
K^{2}A_{\mu }^{\prime }-m_{\gamma }^{2}A_{\mu }^{\prime }-\Pi _{\mu \nu
}\left( K\right) A^{\prime \nu }=0,\ \ \ \ K^{\mu }A_{\mu }^{\prime }=0
\end{equation}

   To specify the components of the polarization tensor, we select a basis
constructed from the following orthogonal four-vectors $\ $ $h^{\mu }\equiv
\left( \omega ,{\bf k}\right) /\sqrt{K^{2}},\ \ \ \ \ l^{\mu
}\equiv \left( k,\omega {\bf n}\right) /\sqrt{K^{2}},$ where the
space-like unit vector ${\bf n}={\bf k}/k$ , $k=|{\bf k}|$ is directed
along the electromagnetic wave vector ${\bf k}$. Thus, the longitudinal
basis tensor can be chosen as $L^{\rho \mu }\equiv -l^{\rho }l^{\mu }$. The
transverse (with respect to ${\bf k}$) components of $\Pi ^{\,\rho \mu }$
have a tensor structure proportional to the tensor $T^{\rho \mu }\equiv
\left( g^{\rho \mu }-h^{\rho }h^{\mu }+l^{\rho }l^{\mu }\right) $ , where $
g^{\rho \mu }={\rm diag}(1,-1,-1,-1)$ is the signature tensor. In
this basis, the polarization tensor has the following form 
\begin{equation}
\Pi ^{\,\rho \mu }\left( K\right) =\pi _{l}\left( K\right) L^{\rho \mu }+\pi
_{t}\left( K\right) T^{\rho \mu },  \label{Pi}
\end{equation}
In the case of a strongly degenerate ultrarelativistic electron plasma $
\left( v_{F}\simeq 1\right) $, the longitudinal and transverse polarization
functions are\footnote{
Our Eq. (\ref{Pil}) differs from Eq. (A39) of the Ref.\cite{BS93} by an
extra factor $\left( \omega ^{2}/k^{2}-1\right) $ because our basis $l^{\mu }
$, $h^{\mu }$ is different from that used by Braaten and Segel. All
components of the complete tensor (\ref{Pi}) identically coincide with that
obtained in \cite{BS93} for the degenerate ultrarelativistic case.} \cite
{BS93}: 
\begin{equation}
\pi _{l}=\frac{1}{D_{e}^{2}}\left( 1-\frac{\omega ^{2}}{k^{2}}\right) \left(
1-\frac{\omega }{2k}\ln \frac{\omega +k}{\omega -k}\right) ,  \label{Pil}
\end{equation}
\begin{equation}
\ \ \ \pi _{t}=\frac{3}{2}\omega _{pe}^{2}\left( 1+\left( \frac{\omega
^{2}}{k^{2}}-1\right) \left( 1-\frac{\omega }{2k}\ln \frac{\omega +k}{\omega
-k}\right) \right) .  \label{Pit}
\end{equation}
The electron plasma frequency and the Debye screening distance are defined
as 
\begin{equation}
\omega _{pe}^{2}=\frac{4}{3\pi }e^{2}\mu _{e}^{2}=\frac{4\pi n_{e}e^{2}}{\mu
_{e}},\ \ \ \ \ \ \ \ \ \ \ \ \ \frac{1}{D_{e}^{2}}=3\omega _{pe}^{2},
\end{equation}
with $\mu _{e}$ and $n_{e}$ being, respectively, the chemical potential and
\ the number density of electrons. The axial polarization tensor is given by 
\begin{equation}
\Pi _{5}^{\,\rho \mu }\left( K\right) =\pi _{A}\left( K\right) ih_{\lambda
}\epsilon ^{\rho \mu \lambda 0},\ \ \ \ \ \ \ \ \ \ \ \pi _{A}\left(
K\right) =\frac{\sqrt{K^{2}}}{2\mu _{e}}\pi _{l}\left( K\right) .
\end{equation}
The total energy of neutrino-pairs is $\omega \sim \Delta \ll \mu _{e}$.
When $\omega \geq k$, polarization functions differ in the order of
magnitude, namely, $\pi _{A}\sim \pi _{l,t}\Delta /\mu _{e}$. Therefore, the
axial polarization of the medium can be neglected.

   In the absence of the electron current, the propagator of a massive photon
is of the form 
\begin{equation}
D_{0}^{\mu \nu }=\frac{4\pi }{K^{2}-m_{\gamma }^{2}}\left( g^{\mu \nu }- 
\frac{K^{\mu }K^{\nu }}{m_{\gamma }^{2}}\right)  \label{D0}
\end{equation}
The in-medium photon propagator, taking into account the induced electron
current, can be found with the aid of the Dyson's equation. The solution is: 
\begin{equation}
D_{\lambda \rho }\left( K\right) =\frac{4\pi }{K^{2}-m_{\gamma }^{2}-\pi
_{l} }L_{\lambda \rho }+\frac{4\pi }{K^{2}-m_{\gamma }^{2}-\pi _{t}}
T_{\lambda \rho }-\frac{4\pi }{m_{\gamma }^{2}}h_{\lambda }h_{\rho }.
\label{D}
\end{equation}
Due to conservation and gauge invariance of the electron current $h_{\lambda
}\Pi ^{\,\lambda\mu }=\Pi ^{\,\lambda \mu }h_{\mu }=0$, the last term of Eq.(
\ref{D}) does not contribute to the effective weak current of a proton in
the medium.

   Insertion of (\ref{Pi}) and (\ref{D}) into (\ref{Jp}) with the above
approximation yields the effective weak current of a proton in the medium. 
\begin{equation}
J^{\mu }=\bar{\psi}\gamma ^{\mu }(C_{V}-C_{A}\gamma _{5})\psi -c_{V}\left( 
\bar{\psi}\gamma _{\lambda }\psi \right) \left( \frac{\pi _{l}L^{\lambda \mu
}}{\omega ^{2}-k^{2}-m_{\gamma }^{2}-\pi _{l}}+\frac{\pi _{t}T^{\lambda \mu }
}{\omega ^{2}-k^{2}-m_{\gamma }^{2}-\pi _{t}}\right)   \label{Jef}
\end{equation}
The poles of this expression give the eigen photon modes in the medium. In
the case of $K^{2}>0$ the eigen mode frequency $\omega \left( k\right) $ \
is larger than the plasma frequency of electrons. At temperatures $
T<T_{c}\ll \omega _{pe}$, the number of such excited oscillations in the
medium is exponentially suppressed, therefore contributions from the poles
can be neglected.

   The contributions of longitudinal and transverse virtual photons to the
effective weak current (\ref{Jef}) are proportional to the following factors 
\begin{equation}
F_{l}\left( k,\omega \right) \equiv \frac{\pi _{l}}{\omega
^{2}-k^{2}-m_{\gamma }^{2}-\pi _{l}}=-\frac{\varphi \left( x\right) }{
D_{e}^{2}\omega ^{2}\left( 1-x^{2}\right) -m_{\gamma
}^{2}D_{e}^{2}\,+\varphi \left( x\right) }
\end{equation}
\begin{equation}
F_{t}\left( k,\omega \right) \equiv \frac{\pi _{t}}{\omega
^{2}-k^{2}-m_{\gamma }^{2}-\pi _{t}}=\frac{1+\varphi \left( x\right) }{
2D_{e}^{2}\omega ^{2}\left( 1-x^{2}\right) -m_{\gamma
}^{2}D_{e}^{2}\,-\left( 1+\varphi \left( x\right) \right) }
\end{equation}
with $x=k/\omega $ and 
\begin{equation}
\varphi \left( x\right) =x^{-2}\left( 1-x^{2}\right) \left( 1-\frac{1}{2x}
\ln \frac{1+x}{1-x}\right) ,\ \ \ \ \ \ \ \ \ \ \ \ m_{\gamma }^{2}D_{e}^{2}=
\frac{\mu _{e}}{3M_{p}^{\ast }}
\end{equation}
The latter equality takes into account local neutrality $N_{p}=n_{e}$ of the
medium\footnote{
Strictly speaking, this equality is valid only at zero temperature, when all
protons are paired.}.

   The neutrino wave-length is much larger than the Debye screening distance,
therefore, $D_{e}^{2}\omega ^{2}\left( 1-x^{2}\right) \lesssim 4\Delta
^{2}D_{e}^{2}\left( 1-x^{2}\right) \ll \varphi \left( x\right) $. Neglecting
this small term we obtain 
\begin{equation}
F_{l}\left( x\right) =-\frac{\varphi \left( x\right) }{\varphi \left(
x\right) -m_{\gamma }^{2}D_{e}^{2}},\ \ \ \ \ \ \ \ \ \ \ \ F_{t}\left(
x\right) =-\frac{1+\varphi \left( x\right) }{1+\varphi \left( x\right)
+m_{\gamma }^{2}D_{e}^{2}}.
\end{equation}
Due to the electron contribution the effective proton interaction with the
neutrino field has a complicated form. According to Eq.(\ref{Jef}), the
vector part of the effective proton weak current is substantially modified
by polarization of the medium. Assuming the proton effective mass to be $
M_{p}^{\ast }\simeq 0.8M_{p}$, functions $F_{l}\left( x\right) $, $
F_{t}\left( x\right) $ are plotted in the Fig. 2 for typical values of the
electron chemical potential $\mu _{e}\simeq 100$ MeV and $\mu _{e}\simeq 200$
MeV.

   For the moment, we shall estimate the contribution of the medium
polarization to the neutrino emissivity caused by formation and breaking of
proton Cooper pairs and leave the complete calculation for a future work. To
roughly estimate the contribution of collective effects, we take advantage
of the fact that $F_{l}(x)$ and $F_{t}(x)$ are slowly varying functions, and
replace them in Eq.(\ref{Jef}) by a constant value $-0.8$. With this
simplification we obtain 
\begin{equation}
J^{\mu }\sim \bar{\psi}\gamma ^{\mu }(C_{V}-C_{A}\gamma _{5})\psi
+0.8c_{V}\,\left( \bar{\psi}\gamma _{\lambda }\psi \right) \left( L^{\lambda
\mu }+T^{\lambda \mu }\right) .  \label{JSR}
\end{equation}
Taking into account the identity $L^{\lambda \mu }+T^{\lambda \mu }\equiv
g^{\lambda \mu }-h^{\lambda }h^{\mu },$ and gauge invariance of the
electromagnetic current of protons $\left( \bar{\psi}\gamma _{\lambda }\psi
\right) _{fi}K^{\lambda }=0,$ we can reduce the effective proton weak
current (\ref{JSR}) as follows 
\begin{equation}
J^{\mu }\sim \bar{\psi}\gamma ^{\mu }(C_{V}+0.8c_{V}-C_{A}\gamma _{5})\psi .
\end{equation}
Thus, the effective vector weak coupling of in-medium protons is of the
order of $C_{V}+0.8c_{V}$ , in contrast to $C_{V}=\frac{1}{2}-2\sin
^{2}\theta _{W}\simeq 0.04$ for bare protons in vacuum. If protons interact
with electron neutrinos then $C_{V}+0.8c_{V}\simeq 0\allowbreak
.\,\allowbreak 8$ , while for muon and tau neutrinos $C_{V}+0.8c_{V}^{\prime
}\simeq 0\allowbreak .008$. The total spin of a Cooper pair in the $
^{1}S_{0} $ state is zero therefore, the contribution to $\nu \bar{\nu}$
emissivity comes only from the vector weak interaction. In previous
calculations \cite{FRS76}, \cite{YKL98} the $\nu \bar{\nu}$ emissivity was
proportional to $C_{V}^{2}$. Taking into account the collective effects, one
has to replace $C_{V}$ by $C_{V}+0.8c_{V}$ for electron neutrinos and by $
C_{V}+0.8c_{V}^{\prime }$ for muon and tau neutrinos. Thus, the $\nu \bar{\nu
}$ emissivity is enhanced about two orders of magnitude: 
\begin{equation}
\frac{Q}{Q_{0}}\sim \frac{\left( C_{V}+0.8c_{V}\right) ^{2}+2\left(
C_{V}+0.8c_{V}^{\prime }\right) ^{2}}{3C_{V}^{2}}\sim \allowbreak 133.
\label{wRatio}
\end{equation}
This ratio demonstrates the large importance of collective processes in the
core of a superconducting neutron star. As $C_{V}\ll c_{V}$, the leading
contribution to $Q$ comes from the weak current of electrons which is
electromagnetically induced by the quantum transition of the initial proton.

   The neutrino emissivity due to Cooper pairing of protons, as well as that
for neutron pairing is proportional to the Fermi momentum of the relevant
particles, i.e. it only weakly $\left( \sim n_{p,n}^{1/3}\right) $ depends
explicitly on the partial number density. Besides that, both the superfluid
and the superconducting energy gaps, calculated theoretically, sensitively
depend on the model of strong interactions and vary in the range from ten
keV to some MeV \cite{Tam70}, \cite{AO85},\cite{B92}, \cite{Ttam93}. Thus,
though the number density of protons is much smaller than that of neutrons,
their partial contribution to the total energy losses is very sensitive to
the ratio between the energy gaps of neutrons and protons. It can be
dominant at some cooling stage, depending on temperature, $T_{cn}$, $T_{cp}$
, and density. Therefore, the neutrino radiation, caused by a
singlet-pairing of protons, should be taken into consideration in the
scenario of cooling of superconducting neutron stars at the same level, as
other basic processes of neutrino production.

   We have shown in the present paper that the effective vector weak current of
in-medium protons is strongly modified by ambient electrons. Their weak
current, electromagnetically induced by the proton quantum transition,
enhances neutrino production due to Cooper pairing of protons by two orders
of magnitude. An analogous conclusion can be made about neutrino-pair
production due to Cooper pairing of $\Xi ^{-}$ hyperons. Though the vector
constant of the weak coupling for $\Xi ^{-}$ hyperons is small relatively to
those for neutrons and $\Sigma ^{\pm }$, the interaction of $\Xi ^{-}$ with
neutrinos is enhanced in the medium by ambient electrons.

\acknowledgments
This work was supported in part by Spanish Grant DGES PB97-
1432, and the Russian Foundation for Fundamental Research Grant 
00-02-16271.

\vskip 0.3cm
\psfig{file=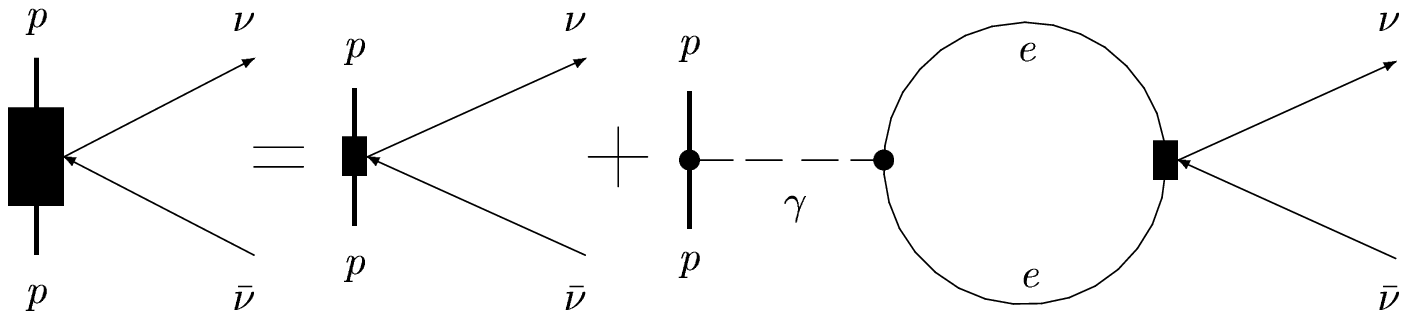}
Fig 1. Feynman graphs contributing to the effective proton interaction
with a neutrino field in the medium. The first diagram presents the
vacuum-like weak interaction. The second diagram, with the electron loop,
describes the interaction via an intermediate virtual photon, 
shown by the dashed line.
\vskip 0.3cm

$\allowbreak $

\vskip 0.3cm
\psfig{file=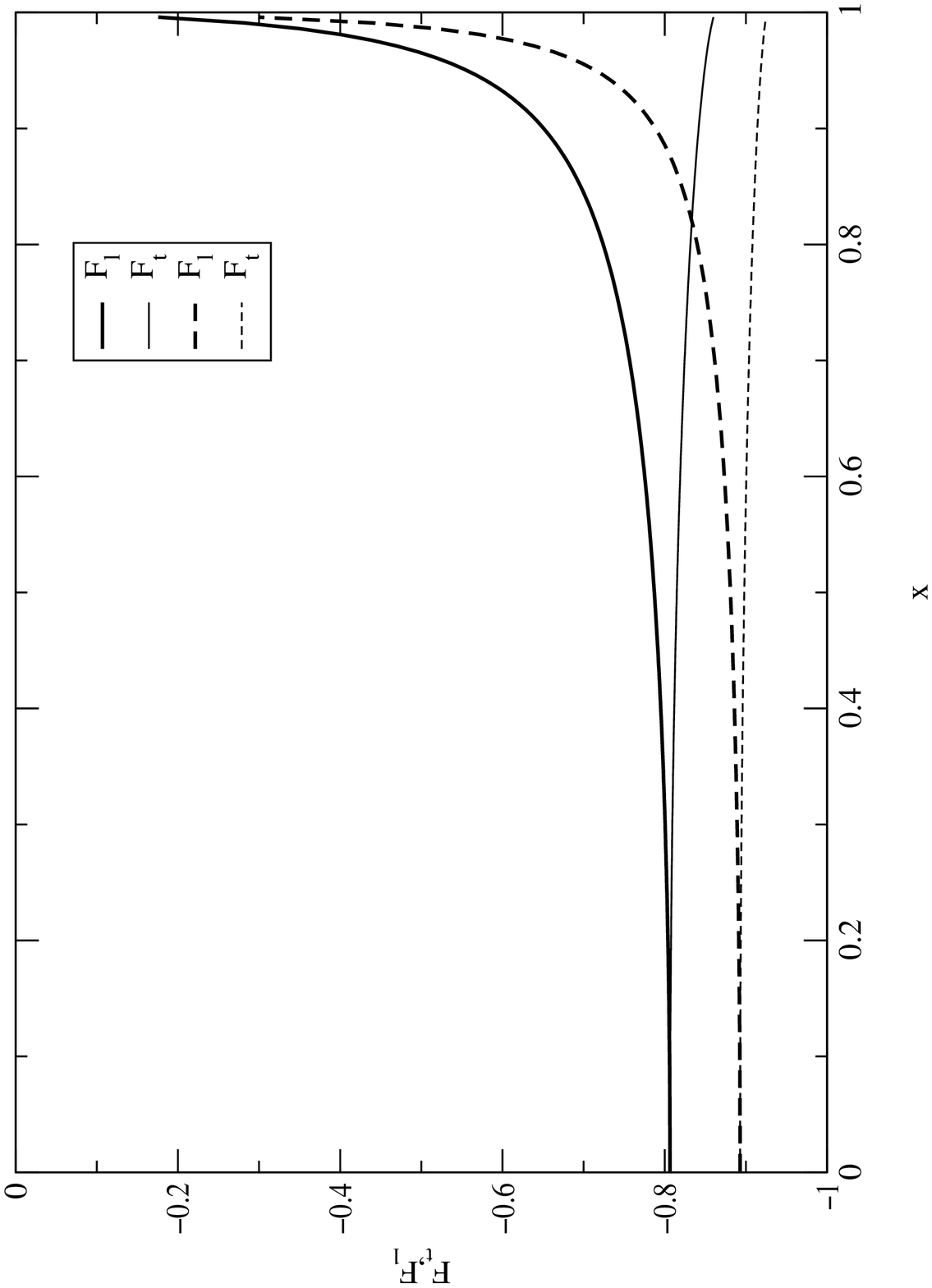}
Fig 2. Functions $F_{l}\left( x\right)$, $F_{t}\left( x\right)$
are plotted for two cases of the
electron chemical potential $\mu _{e}\simeq 100$ MeV (deshed lines) 
and $\mu _{e}\simeq 200$ MeV (solid lines). The proton effective mass  
$M_{p}^{\ast } = 0.8M_{p}$
\vskip 0.3cm

\end{document}